\documentstyle{article}
\font\tenbf=cmbx10
\font\tenrm=cmr10
\font\tenit=cmti10
\font\elevenbf=cmbx10 scaled\magstep 1
\font\elevenrm=cmr10 scaled\magstep 1
\font\elevenit=cmti10 scaled\magstep 1

\font\ninerm=cmr9

\renewenvironment{thebibliography}[1]
 { \elevenrm
   \begin{list}{\arabic{enumi}.}
    {\usecounter{enumi} \setlength{\parsep}{0pt}
     \setlength{\itemsep}{3pt} \settowidth{\labelwidth}{#1.}
     \sloppy
    }}{\end{list}}

\def\be{\begin{equation}}
\def\ee{\end{equation}}
\def\ba{\begin{eqnarray}}
\def\ea{\end{eqnarray}}
\def\bZ{{\bf Z}}
\def\ket#1{\vert #1\rangle}

\begin{document}
\begin{flushright}CTP\#2183\end{flushright}
\vglue 0.5cm
\begin{center}{{\tenbf FRACTIONAL SUPERSTRING\\
               \vglue 10pt TREE SCATTERING AMPLITUDES\footnote{
\ninerm\baselineskip=11pt Talk presented at the 26th Erice workshop:
``From Superstrings to Supergravity''.}\\}
\vglue 1.0cm
{\tenrm PHILIP C. ARGYRES\footnote{\ninerm\baselineskip=11pt
Visiting the Center for Theoretical Physics, M.I.T., Cambridge,
MA 02139 until June 1993.
E-mail address: pca@strange.tn.cornell.edu.} \\}
\baselineskip=13pt
{\tenit Newman Laboratory,
Cornell University\\}
\baselineskip=12pt
{\tenit Ithaca, NY 14853, USA\\}
\vglue 0.8cm
{\tenrm ABSTRACT}}
\end{center}
\vglue 0.3cm
{\rightskip=3pc
 \leftskip=3pc
 \tenrm\baselineskip=12pt
 \noindent
The spin-4/3 fractional superstring is characterized by a chiral
algebra involving a spin-4/3 current on the
world-sheet in addition to the energy-momentum tensor.
These currents generate physical state conditions on the
fractional superstring Fock space.  Scattering amplitudes of
these physical states are described which satisfy both
spurious state decoupling and cyclic symmetry (duality).
Examples of such amplitudes are calculated using an explicit
$c=5$ realization of the spin-4/3 current algebra.
This representation has three flat coordinate boson
fields and a global SO(2,1) Lorentz symmetry, permitting
a particle interpretation of the amplitudes.
\vglue 0.6cm}
%===Text=======================================================
\vglue 0.4cm
\baselineskip=14pt
\elevenrm
String theories are characterized by the
local symmetries of two-dim\-en\-sion\-al field theories
on the string world-sheet.
The bosonic string is invariant under diffeomorphisms
and local Weyl rescalings on the world-sheet;
whereas the superstring is characterized by
a locally supersymmetric version of these
symmetries.  It is natural to ask whether other
symmetries on the world-sheet can give rise
to consistent string theories.
Recently, a proposal for a large class of
new string theories, called fractional superstrings, was
advanced.$^{1}$  Since fractional-spin
fields exist in two-dimensional theories, one can
imagine new local symmetries on the world-sheet
involving fractional-spin currents (replacing the
spin-3/2 supercurrent of the superstring).  Evidence
has been presented$^{2,3,4}$ that fractional superstrings
with spin 4/3, 6/5, and 10/9 currents on the world-sheet
have potentially interesting phenomenologies in 6, 4 and 3
space-time dimensions, respectively.

This talk will focus exclusively on the spin-4/3 fractional
superstring and its tree scattering amplitudes.
In the course of the discussion I will simply state many of the
properties of the representation theory of the spin-4/3
current algebra on the string world-sheet;  for proofs
and more details, see Ref.~5.
Classically, the spin-4/3 algebra is the constraint
algebra arising from gauge-fixing the local world-sheet
symmetry.  Quantum mechanically, the constraints generate
physical state conditions which pick out the propagating
degrees of freedom from the larger string state space.
Although the classical world-sheet gauge symmetry giving
rise to a spin-4/3 constraint algebra is not understood
at present, we can make progress by taking the constraint
algebra itself as a starting point, and checking the consistency
of the resulting string theory by constructing unitary scattering
amplitudes for the physical states.
\vglue 0.6cm
%===============================================================
{\elevenbf\noindent 1. The spin-4/3 fractional superconformal
chiral algebra}
\vglue 0.4cm
Before describing the scattering amplitudes, we must first define
in more detail the spin-4/3 fractional superstring.  The motivating idea
behind the construction of this string is to replace the
spin-1/2 world-sheet fermions $\psi^\mu$ appearing in the
ten-dimensional
superstring with spin-1/3 world-sheet fields $\epsilon^\mu$.
The world-sheet supersymmetry
of the superstring is then replaced with a
world-sheet ``fractional supersymmetry''
which relates coordinate boson fields
$X^\mu$ not to fermions but rather to
the field $\epsilon^\mu$.  The fractional
supersymmetry is generated by a
generalization of the supercurrent,
a new chiral current $G$ of the form $G(z)\sim\epsilon^\mu
\partial X_\mu+\cdots$, whose conformal dimension is 4/3.

The fractional current, $G(z)$,
and the energy-momentum tensor, $T(z)$,
together generate the
fractional superconformal (FSC) chiral algebra,
encoded in the singular terms of the operator product expansions:
\ba\label{truefss}
 T(z)T(w)&=&{1\over(z-w)^4}\left\{{c\over2}+2(z-w)^2T(w)
  +(z-w)^3\partial T(w)\right\}~,\nonumber\\
 T(z)G(w)&=&{1\over(z-w)^2}\left\{{4\over3}G(w)
  +(z-w)\partial G(w)\right\}~,\nonumber\\
 G(z)G(w)&=&{1\over(z-w)^{8/3}}\left\{{3c\over4}+
   2(z-w)^2 T(w)\right\}\nonumber\\
 &&\mbox{}+{\lambda\over(z-w)^{4/3}}
   \left\{G(w)+{1\over2}(z-w)\partial G(w)\right\}.
\ea
The first OPE just states that $T(z)$ obeys the Virasoro algebra,
while the second implies that $G(z)$
is a dimension-4/3 chiral primary field.
We will take this algebra as the constraint
algebra generating the physical state conditions for
the spin-4/3 fractional string.
Although one can imagine other chiral algebras
involving dimension-4/3 currents, we have chosen the
above FSC algebra to define the spin-4/3 fractional superstring
because this algebra is known to have a sensible representation
theory.$^{6,7,8}$  In particular, it is
known that associativity fixes $\lambda$ as a function of the
central charge $c$:
\be\label{lamb}
  \lambda^2={8-c\over6}.
\ee

An important feature of the FSC algebra is the appearance of
cuts in the $GG$ OPE.  Since there
are two different cuts on the right hand side,
upon continuation of a correlation function involving
$G(z)G(w)$ along a contour interchanging $z$ and $w$
it is not consistent for the correlator to pick up a
simple phase.  This situation is described by saying
that the current $G$ satisfies ``nonabelian braid relations.''
For such currents, their OPEs alone do not define
the chiral algebra; they must be supplemented
by the current braid relations.$^{5}$
These braid relations can be described in a simple
way.  Since only two fractional cuts appear in the $GG$ OPE,
one can split the $G$ current into two pieces,
$G(z)=G^+(z)+G^-(z)$,
which satisfy abelian braid relations:$^{7}$
\ba\label{ftalg}
  G^{\pm}(z)G^{\pm}(w)&=&{\lambda\over(z-w)^{4/3}}
   \left\{G^{\mp}(w)+{1\over2}(z-w)\partial
G^{\mp}(w)\right\},\nonumber\\
  G^{\pm}(z)G^{\mp}(w)&=&{1\over(z-w)^{8/3}}\left\{{3c\over8}+
   (z-w)^2T(w)\right\}.
\ea
Under interchange of $z$ and $w$ (along a prescribed path, say
a counterclockwise switch) the only consistent phase that
$G^+$ or $G^-$ can pick up with itself is $e^{4i\pi/3}$.
The phase that develops upon interchange of $G^+$ with $G^-$
can be taken to be $e^{2i\pi/3}$.
This serves to define the braid relations satisfied
by the full FSC current $G(z)$.  It is worth emphasizing
that the split algebra currents $G^\pm$ do {\elevenit not}\/ generate
the physical state conditions for the spin-4/3 superstring; only
their combination $G=G^++G^-$ does.  The split algebra
of Eq.~(\ref{ftalg}) is only introduced as a technical tool
to describe the braid relations of the fractional supercurrent
$G(z)$.

The specification of the FSC algebra, Eq.~(\ref{truefss}), along with
the braid relations of $G$ defines the spin-4/3 fractional
superstring.  The rest of this talk will outline the first step
towards showing that this string in fact exists, by
constructing tree scattering amplitudes which are consistent
with the physical state conditions following from the FSC
algebra.  We will see by explicit calculation in an example
that these scattering amplitudes have a sensible
space-time interpretation.
\vglue 0.6cm
%===============================================================
{\elevenbf\noindent 2. Physical state conditions and FSC
highest-weight modules}
\vglue 0.4cm
The first step in constructing tree scattering
amplitudes is to derive the algebra of the
constraints from the FSC algebra.  The physical
state conditions are imposed by the requirement that
the positive (annihilation) modes of the FSC currents
vanish when acting on physical states.  Deriving the
constraint algebra is thus equivalent to deriving from
the OPEs and braid relations the
commutation relations satisfied by the modes of the
currents.  This is a technically complicated step
the results of which I will simply state; for a
fuller derivation see Ref.~5.

The split algebra, Eq.~(\ref{ftalg}),
was studied by Zamolodchikov and Fateev,$^{7}$ who noted
a $\bZ_3$ symmetry which organizes its representation theory.
In particular,
the currents $G^{+}$ and $G^{-}$ can be assigned
${\bf Z}_3$ charges $q=1$ and $-1$,
respectively, while the energy-momentum tensor $T$ (as well
as the identity) have charge $q=0$.
It is natural to assume that, since the split algebra
is supposed to be an organizing symmetry of our theory, all the fields
in a representation have definite $\bZ_3$ charges.\footnote{
\ninerm\baselineskip=11pt
One can also introduce fields which are double-valued with respect
to the $G^\pm$ currents;$^{7}$ this corresponds to enlarging
the symmetry
of the split algebra from $\bZ_3$ to $S_3$, its full automorphism
group.  The fields in this double-valued sector play a role in
the spin-4/3 string analogous to that played by the Ramond sector
fields in the superstring.$^{5}$}

Since the split algebra is abelianly braided, the
arguments of Ref.~6 can be directly applied to derive the
mode expansions and generalized commutation relations
following from Eq.~(\ref{ftalg}).
The mode expansions of $T$, $G^{+}$ and $G^{-}$ are
defined by
  \ba\label{Jmode}
  T(z)\chi_q(0)&=&\sum_n z^{-n-2} L_n\chi_q(0)~,\nonumber\\
  G^{\pm}(z)\chi_q(0)&=&\sum_n z^{n\mp q/3}
   G^{\pm}_{-1-n-(1\mp q)/3}\chi_q(0)~,
  \ea
where $\chi_q$ is an arbitrary state with ${\bf Z}_3$ charge $q$.
This means, in particular, that the $L_n$'s have integral
moding, while the $G_r$'s are moded by one-third integers.
The mode expansion for the full fractional superconformal
current can be built from the split algebra pieces: $G_r=G^+_r+G^-_r$.
The modes of the $G^\pm$ currents and the energy-momentum tensor
$T$ satisfy the commutation relations
\ba\label{Jalgcom}
 \left[L_m,L_n\right]&=&(m-n)L_{m+n}
 +{c\over12}(m^3-m)\delta_{m+n}~,\nonumber\\
 \left[L_m,G^\pm_r\right]&=&
 \left({m\over3}-r\right)G^\pm_{m+r}~,
\ea
and the generalized commutation relations$^{7}$
\ba\label{Jalg}
 &&\sum_{\ell=0}^\infty c^{(-2/3)}_\ell\left[
  G^\pm_{\pm{q\over3}+n-\ell}G^\pm_{{2\pm q\over3}+m+\ell}-
  G^\pm_{\pm{q\over3}+m-\ell}G^\pm_{{2\pm q\over3}+n+\ell}\right]
  ~=~{\lambda\over2} (n-m)G^\mp_{{2\pm2q\over3}+n+m}~,\nonumber\\
 &&\sum_{\ell=0}^\infty c^{(-1/3)}_\ell\left[
  G^{+}_{{1+q\over3}+n-\ell}G^{-}_{-{1+q\over3}+m+\ell}+
  G^{-}_{-{2+q\over3}+m-\ell}G^{+}_{{2+q\over3}+n+\ell}\right]
 ~=~\nonumber\\
 &&\qquad\qquad\qquad\qquad\qquad\qquad
  L_{n+m}+{3c\over16} \left(n+1+{q\over3}\right)
  \left(n+{q\over3}\right)\delta_{n+m}~,
\ea
when acting on a state with ${\bf Z}_3$ charge $q$. The
$c^{(\alpha)}_\ell=(-1)^\ell{\alpha\choose\ell}$ are
fractional binomial coefficients.
Because of the infinite sum on the left hand side, the
mode algebra in Eq.~(\ref{Jalg}) is not a graded Lie
algebra, but a new algebraic structure on the string world-sheet.

The physical state conditions require that a physical state
$\ket{\phi}$ be annihilated
by the positive modes of $T$ and $G$, and be eigenstates of
their zero modes:
  \ba\label{psc}
  (L_n-v\delta_{n,0})\ket{\phi}&=&0~,\quad 0\le n\in\bZ\ ,\nonumber\\
  (G_r-\beta\delta_{r,0})\ket{\phi}&=&0~,\quad 0\le r\in\bZ/3\ .
  \ea
Here $v$ and $\beta$ are ``intercepts'', normal ordering
constants in the definitions of $T$ and $G$.
All the positively-moded constraints can be generated from
those of the set $\{L_1,L_2,G_{1/3},G_{2/3},G_1,G_{4/3}\}$.
A state $\ket{s}$ obeying the zero-mode conditions in Eq.~(\ref{psc})
is called a spurious state if it is orthogonal to all
physical states.  Such a state can be written as
\be\label{spur}
\ket{s}=\sum_{n>0}L_{-n}\ket{\chi_n}+\sum_{r>0}G_{-r}\ket{\psi_r}\ ,
\ee
in terms of some other states $\ket{\chi_n}$ and $\ket{\psi_r}$.
All states not satisfying the physical state conditions must
have a spurious component.  A physical state can itself be spurious,
in which case it is a null state, and should decouple from all
scattering
amplitudes.  Thus, the decoupling of all spurious states in the
scattering amplitudes of physical states is a test of whether
Eq.~(\ref{Jalg}) is a sensible constraint algebra.

The physical state conditions imply that the physical
state $\ket{\phi}$ is a highest weight state of the FSC
algebra with highest weight $v$.
The properties of the highest weight modules can be derived
from the commutation relations Eqs.~(\ref{Jalgcom}--\ref{Jalg}),
and it turns out that they fall into two classes.$^{7}$
The highest weight state of an S
(singlet) module has $\bZ_3$ charge $q=0$, the
allowed modings of the fractional current acting on it are
$G_{n-(1/3)}$, and its descendants have dimensions $v+n$ and
$v+{1\over3}+n$ where $n$ is an integer.  The highest weight
state of a D (doublet) module is doubly degenerate in
the split algebra with $\bZ_3$ charges $q=\pm1$, allowed
current modings $G_{n}$ and $G_{n-(2/3)}$, and descendants
of dimensions $v+n$ and $v+{2\over3}+n$.
In the D sector, associativity fixes the $G_0$ intercept
$\beta$ in terms of $v$ and $c$.

We will see later,
by way of an example, that physical states in the S sector
are space-time bosons, and include a tachyon; those
in the D sector are also space-time bosons, but with the lightest
state a massless vector particle (or graviton for a closed
string).\footnote{\ninerm\baselineskip=11pt
As mentioned earlier, there is also a third
sector consisting of highest weight modules double-valued with
respect to the fractional currents, called the R sector.
The appropriate moding of the currents in this sector is
half-integral, and the algebra obeyed by the
modes is different from Eq.~(\ref{Jalg}).  The physical states
in this sector correspond to space-time fermions.$^{5}$}
Henceforth we focus on the properties of the D modules, since it is
only for the D-sector states that we can construct scattering
amplitudes.

The main properties of D modules are summarized by the OPEs of the
current $G$ with the highest weight vertex operator $W_D(z)$ of
dimension-$v$ and its first descendant operator $V_D(z)$ of
dimension $v+{2\over3}$:$^{5}$
\ba\label{Drep}
G(z)W_D&=&\left(\sqrt{v-{c\over24}}\right){W_D\over z^{4/3}}+
 {V_D\over z^{2/3}}+\ldots\nonumber\\
G(z)V_D&=&\left(v+{c\over12}+{1\over12}\sqrt{(8-c)(24v-c)}\right)
 {1\over z^2} \left\{W_D + {z\over3v} \partial W_D\right\}\nonumber\\
&&\qquad\mbox{}+\left({\sqrt{24v-c}-\sqrt{8-c}\over36v}\right)
 {\widetilde W_D\over z} +\ldots
\ea
For ease of writing, we have inserted the D-module vertex operators
$W_D$ {\elevenit etc.}\ at the origin of the complex plane and
have dropped their arguments.   The first OPE fixes the
normalization of the $V_D$ vertex, while the second one
introduces the new Virasoro (though not FSC) primary
$\widetilde W_D=(L_{-1} - 3\sqrt{6}vG_{-1})W_D$
of dimension $v+1$.  Note that the fractional current is
single-valued with respect to $V_D$, but is nonabelianly
braided with respect to $W_D$.
These OPEs play an important role in the development
of consistent tree scattering amplitudes for D-sector physical
states.
\vglue0.6cm
%===============================================================
{\elevenbf\noindent 3. Tree scattering amplitudes and
spurious state decoupling}
\vglue 0.4cm
In what follows, we will construct open fractional
superstring tree scattering amplitudes.
Closed string scattering amplitudes at tree level are easily
formed by combining two open string amplitudes using a
level-matching condition for left- and right-movers.
The construction we use will be closely analogous to that of open
Ramond-Neveu-Schwarz superstring tree amplitudes in the ``old
covariant'' formalism.  (For a detailed explanation of this
formalism, see, {\elevenit e.g.}, Chapter 7 of Ref.~9.)

The world-sheet in an open string tree scattering process is
conformally equivalent to a unit disc with vertex operators
$V(x)$ representing the asymptotic scattering states
inserted at points on the boundary.  Since we must be able to
integrate these vertex operators over their insertion
positions, they must be dimension-one operators in the
two-dimensional world-sheet theory.
We can conformally map the disk to the complex upper half-plane,
fixing the positions of three of the vertex insertions at
$\infty$, 1, and 0 on the real axis, with the remaining insertions
at points $1<x_i<\infty$.  This picture is suitable for writing
the string amplitude as a correlator of operators in the open
string Fock space with a radial-ordering prescription:
\be\label{Aone}
{\cal A}_N= \int {dx_3\cdots dx_{N-1}\over x_3\cdots x_{N-1}}
\langle  V_N| V_{N-1}(x_{N-1})\cdots V_2(1)\ket{ V_1}\ ,
\ee
where the ``in'' and ``out'' states are the insertions at $x_1=0$ and
$x_N=\infty$, and the integration is over all $x_i$ preserving
the order $1<x_3<\cdots<x_{N-1}<\infty$.

{}From our previous
mapping to the disk, it is clear that ${\cal A}_N$ should be
invariant under cyclic permutations of the vertex ordering.
This means that after passing the $V_N$ vertex to the right through
all the other vertices in Eq.~(\ref{Aone}), the value of ${\cal A}_N$
must be unchanged.  This condition on the braiding of the
vertex operators can be satisfied if two such operators commute
when zero space-time momentum is flowing into either vertex.
The representation theory of the spin-4/3 split algebra
implies that commuting operators can only have $\bZ_3$ charge $q=0$.
Thus, in particular, the $V_D$ (but not the $W_D$) operator
in the D sector is a potential candidate for the vertex insertions
in Eq.~(\ref{Aone}). For this also to be a dimension-one
operator implies that the D-sector intercept must
be  $v=1/3$.

A vertex insertion at $x$ can be rewritten
as $ V(x) = x^{L_0} V(1) x^{-L_0}$, and the
positions of the insertions explicitly integrated over to give
the amplitude in the form
\be\label{Aprop}
{\cal A}_N= \langle V_N| V_{N-1}(1)
\widetilde\Delta\ldots\widetilde\Delta
 V_2(1)\ket{ V_1}
\ee
where the propagator is $\widetilde\Delta=(L_0-1)^{-1}$.
For scattering of D-sector states, the vertices in Eq.~(\ref{Aprop})
correspond to the $V_D$ descendant states in a
FSC highest-weight module.  We can convert Eq.~(\ref{Aprop})
to a different ``picture'' involving the highest-weight states $W_D$
using the general properties of the D modules.
In particular, by taking appropriate fourier components
of the D-sector OPE in Eq.~(\ref{Drep}), it follows that
\be\label{gocom}
{}[G_r,V_D(1)]=\left(L_0+r-{1\over3}\right)W_D(1)
-W_D(1)\left(L_0-{1\over3}\right)
\qquad\qquad{\rm for\ all}\ r\in\bZ/3\ .
\ee
In deriving this formula we have set $v=1/3$ and used
the relation $[L_0,W_D(1)]=(1/3)W_D(1)+\partial W_D(1)$
which is true for any dimension-1/3 Virasoro primary $W_D$.
Note that when $v=1/3$
the dimension $1+v$ descendant $\widetilde W_D$ (which has no
analog in the superstring) decouples from the OPE in Eq.~(\ref{Drep}).
It is this unexpected decoupling at precisely the physical value
of the intercept which allows us to derive Eq.~(\ref{gocom}), and
in turn construct sensible tree scattering amplitudes.

\setcounter{footnote}{0}
We can now replace the ``in'' state $\ket{V_D}$ with a
physical state using $\ket{V_D}=G_{-2/3}\ket{W_D}$ which
follows from the first OPE in Eq.~(\ref{Drep}).  The $G_{-2/3}$
mode can be commuted to the left using Eq.~(\ref{gocom})
as well as the relation
\be\label{propcom}
G_r(L_0-a-r)^{-1}=(L_0-a)^{-1}G_r\ ,
\ee
following from Eq.~(\ref{Jalgcom}).  Acting on the ``out'' state,
$\langle V_D|G_{-2/3}=\langle W_D|$, which is a consequence of the second
OPE in Eq.\ (\ref{Drep}) with $v=1/3$.  The extra insertions coming
from the right-hand side of Eq.~(\ref{gocom}) vanish by
a ``cancelled propagator'' argument, since
setting $r=-2/3$ in Eq.~(\ref{gocom}) gives factors of $L_0-1$ and
$L_0-1/3$ which cancel the propagators to the left and right,
respectively.\footnote{\ninerm\baselineskip=11pt
Tree amplitudes with cancelled propagators are holomorphic in
the Mandelstam invariant of the cancelled propagator channel, and
thus vanish if the amplitudes have Regge asymptotic behavior. We
will see in the next section that they do have this soft high
energy behavior.}
Thus, the final form we find for the amplitude is
\be\label{Apict}
{\cal A}_N=\langle W_N|V_{N-1}(1)\Delta\ldots\Delta
V_2(1)\ket{W_1}\ ,
\ee
where the propagator in this picture is $\Delta=(L_0-{1\over3})^{-1}$.

Now we can investigate the crucial issue of spurious
state decoupling in our amplitudes.  If we start with
physical states defined as highest-weight vectors of
FSC modules, will they scatter only
to other physical states?  For this to be true, only
physical states must contribute to residues of poles in
amplitudes when an internal propagator goes on-shell.
Suppose we fix the external momenta such that some
state $\ket{s}$ in the string Fock space at momentum
$\kappa=k_{M+1}+\cdots+k_N$ is on-shell:
$(L_0-1/3)\ket{s}=0$.  If we factorize the amplitude
in Eq.~(\ref{Apict}) by inserting a sum over a complete set of states of
momentum $\kappa$ at the propagator between $V_{M+1}$
and $V_M$, then the $\ket{s}\langle s|$ term in the
sum will contribute a pole in momentum space.  The
requirement of spurious state decoupling is that if $\ket{s}$
is spurious, its contribution to the residue of the
pole should vanish:
\be\label{spstde}
\langle s|V_M(1)\Delta\cdots\Delta V_2(1)\ket{W_1}=0\ .
\ee
To prove this,
consider one term, $\langle\psi|G_r$ with $r>0$, in the
presentation of $\langle s|$ as a sum of descendant states,
Eq.~(\ref{spur}).  (The $L_n$ descendant pieces can be
shown to decouple by a similar argument.)  The $G_r$ mode
can be commuted to the right in Eq.~(\ref{spstde}) using
Eqs.~(\ref{gocom}--\ref{propcom}).  The
insertions coming from the right-hand side of Eq.~(\ref{gocom})
again vanish by a cancelled
propagator argument.  Finally,
the $G_r$ mode acting on the ``in'' state $\ket{W_1}$ vanishes
by the physical state conditions Eq.~(\ref{psc}), thus
proving spurious state decoupling.

Our prescription for constructing dual N-point
tree amplitudes of D-sector states satisfying spurious state
decoupling can be extended to include one or two S-sector
states by simply replacing the ``in'' and ``out'' $W_D$
states in Eq.~(\ref{Apict}) with S-module physical states $W_S$.
The argument for spurious state decoupling then goes through unchanged.
However, it turns out that there is no appropriate dimension-1
commuting vertex in the S sector to play the role of the $V_D$
vertices.  Thus, we cannot prove cyclic symmetry of the
amplitudes with two S-sector states, nor can we extend the
prescription to include scattering of three or more S-sector states.
Presumably, as in the Ramond sector of the superstring,
this means that there is a nontrivial contribution to
S-sector scattering amplitudes coming from the ``fractional
ghost'' fields on the world-sheet.

In section 6 we will argue that a consistent intercept in the
S sector is $v=1/3$, the same as that of the D sector.  Note,
however, that upon factorizing the D-sector scattering
amplitude in Eq.~(\ref{Apict}) on any propagator, we can never
obtain an S-sector intermediate state.  The reason is simply
that the $W_D$ ``in'' state has $\bZ_3$ charge $q=\pm1$ in
the split algebra, and the $V_D$ vertices have charge $q=0$.
Thus, by conservation of $\bZ_3$ charge, only $q=\mp1$
intermediate states can contribute, whereas the S-sector
physical states $W_S$ have $q=0$.  This selection rule means
that it is consistent at tree level to drop the S sector
altogether, a desirable feature since it will turn out
that the S sector contains tachyons.\footnote{\ninerm\baselineskip=11pt
The considerations of the last two paragraphs also apply
to the R sector: spurious state decoupling works for amplitudes
with just two R-sector vertices, though cyclic symmetry is
not manifest; and D-sector scattering does not produce
R-sector states, though in this case this is interpreted
as a space-time spin-statistics selection rule, not as a means
of decoupling the R sector.}
\vglue 0.6cm
%===============================================================
{\elevenbf\noindent 4. A c=5 free-field representation of the FSC algebra}
\vglue 0.4cm
The scattering amplitude construction we have presented so far
has depended only on general properties of the highest-weight
modules of the FSC algebra.  In this section we will flesh out
this construction with an explicit conformal field theory which
forms a representation of the FSC algebra at central charge $c=5$.
In the next section we will argue that the critical central charge
of the FSC algebra is $c=10$, so the representation presented below
is sub-critical.  As is the case with sub-critical representations
of the bosonic and superstring, tree amplitudes are perfectly
well-behaved.  The restriction to the critical central charge is
expected only to appear once loop amplitudes are included.

The $c=5$ representation is constructed from five free (massless)
scalar fields on the world-sheet.  Three of them, $X^\mu(z)$, $\mu=0,1,2$,
are interpreted as the string coordinate fields, and obey the
standard OPE
\be\label{Xope}
X^\mu(z) X^\nu(w)=-\eta^{\mu\nu}\ln(z-w)\ ,
\ee
where $\eta^{\mu\nu}$ is the three-dimensional Minkowski metric
with signature $(-++)$.  This $X^\mu$ CFT has a global SO(2,1)
Lorentz symmetry.  The remaining two fields, $\varphi^i(z)$, $i=1,2$,
are compactified on a triangular lattice---the su(3) root lattice.
In a basis in which the $\varphi^i$ boundary conditions are
diagonalized, $\varphi^i=\varphi^i+2\pi$, their OPEs read
\be\label{phiope}
\varphi^i(z)\varphi^j(w)=-g^{ij}\ln(z-w)\ ,\qquad
g^{ij}={1\over3}\pmatrix{\hfill 2&-1\cr -1&\hfill 2\cr}\ .
\ee
The vertex operator $V_{\bf m}={\rm exp}\{im_j\varphi^j\}$ has
dimension $\Delta_{\bf m}={1\over2}m_ig^{ij}m_j$ for integer $m_i$.
A triplet of dimension-one fields, $U^\mu=
\{[V_{(2,1)}+V_{(-2,-1)}] ,[ V_{(-1,1)}+V_{(1,-1)}] ,[
V_{(1,2)}+V_{(-1,-2)}]\}$, generates an SO(2,1)$_2$
Kac-Moody algebra,
organizing all the fields in the $\varphi^i$ CFT in SO(2,1)
representations.\footnote{\ninerm\baselineskip=11pt
Appropriate, though standard, cocycles must be added
to the $\varphi^i$ CFT to realize this symmetry.}
Some important fields of low conformal dimension $\Delta$ are:
\ba\label{covflds}
\Delta=1/3:&\epsilon_\mu=\left\{V_{(1,0)},V_{(0,1)},V_{(-1,-1)}\right\}\ ,
\   \epsilon^\dagger_\mu=\left\{V_{(-1,0)},V_{(0,-1)},V_{(1,1)}\right\}&
\nonumber\\
\Delta=1\ \,\ :&W_{\mu\nu}=\left\{{i\over2}\left[V_{(2,1)}-V_{(-2,-1)}\right],
{i\over2}\left[V_{(1,-1)}-V_{(-1,1)}\right],
{i\over2}\left[V_{(-1,-2)}-V_{(1,2)}\right],
i\partial\varphi^j\right\}&\nonumber\\
\Delta=4/3:&s={1\over3}\left[V_{(-2,0)}+V_{(0,-2)}+V_{(2,2)}\right]\ ,
\ s^\dagger={1\over3}\left[V_{(2,0)}+V_{(0,2)}+V_{(-2,-2)}\right]\ .&
\ea
The dimension-1/3 vector fields, $\epsilon_\mu$ and $\epsilon^\dagger_\mu$,
are the analogs of the dimension-1/2 $\psi^\mu$ fields in the flat-space
superstring representation.  The scalars, $s$ and $s^\dagger$, and the
spin-2 (symmetric, traceless) $W_{\mu\nu}$ have no superstring analogs.

A current obeying the FSC algebra Eq.~(\ref{truefss}) is given by
\be\label{Grepd}
G={1\over\sqrt2}\left[(\epsilon_\mu-\epsilon^\dagger_\mu)\partial X^\mu
-{3\over2}(s+s^\dagger)\right]\ .
\ee
It is manifestly an SO(2,1) scalar Virasoro primary dimension-4/3
field.  A splitting of the current satisfying the split algebra
in Eq.~(\ref{ftalg}) exists
with $\bZ_3$ charge assignments $q=1$ for
$\epsilon_\mu$ and $s$, $q=-1$ for
$\epsilon^{\dagger}_\mu$ and $s^{\dagger}$, and $q=0$ for
$X^\mu$, $U^\mu$, and $W^{\mu\nu}$.\footnote{\ninerm\baselineskip=11pt
Fields in the R sector appear upon orbifolding the $\varphi^i$ CFT
by a 180$^\circ$ rotation of its lattice.  The twist fields of
this $\bZ_2$ orbifold transform in the spinor representation of
SO(2,1), and the zero modes of the $\epsilon^\mu$ fields acting on
these states satisfy a three-dimensional Clifford algebra.$^{5}$}

The simplest S-sector vertex is $W_S={\rm exp}\{ik\cdot X\}$.
The physical state conditions, Eq.~(\ref{psc}), applied to $W_S$
imply that $k^2=2v$.  We will show in the next section that a
consistent choice for the S-sector intercept is $v=1/3$, which implies
that $W_S$ is a tachyon.

The simplest D-sector vertex is $W_D=(\zeta\cdot\epsilon+
\zeta^\prime\cdot\epsilon^\dagger){\rm exp}\{ik\cdot X\}$.  The physical
state conditions acting on $W_D$ constrain $\zeta_\mu$ and
$\zeta^\prime_\mu$ to be expressible in terms of a single polarization
vector $\xi_\mu$, such that $k^2=k\cdot\xi=0$, and
\be\label{vectvert}
W_D=\left[\xi^\mu(\epsilon_\mu-\epsilon^\dagger_\mu)
-\varepsilon^{\mu\nu\rho}\xi_\mu k_\nu(\epsilon_\rho
+\epsilon^\dagger_\rho)\right]{\rm e}^{ik\cdot X}\ .
\ee
Here $\varepsilon^{\mu\nu\rho}$ is the completely antisymmetric
tensor in three dimensions.  $W_D$ represents a massless vector
particle.  In a closed string, we could match a left-moving and a
right-moving version of $W_D$ to form the usual graviton, dilaton,
and antisymmetric tensor fields.  The $V_D$ descendant of $W_D$,
defined by the first OPE in Eq.~(\ref{Drep}), is
\be\label{Vvert}
V_D=-\sqrt{2}\Bigl[\xi^\mu\partial X_\mu
-\varepsilon^{\mu\nu\rho}\xi_\mu k_\nu U_\rho
-ik^\mu\varepsilon^{\nu\rho\sigma}\xi_\rho k_\sigma W_{\mu\nu}
\Bigr]{\rm e}^{ik\cdot X}\ .
\ee
Note that upon making a gauge transformation $\delta\xi^\mu\propto
k^\mu$, one finds $\delta V_D\propto\partial({\rm exp}\{ik\cdot X\})$,
a spurious state which decouples by the arguments of the last section.
The operators $\partial X_\mu$, $U_\mu$, and $W_{\mu\nu}$ appearing in
$V_D$ are all single-valued fields on the world-sheet---they have no
cuts in their OPEs with any other field---a
property shared by all operators with $\bZ_3$ charge $q=0$.
Therefore, only simple commutators, as opposed to the generalized
commutators appearing in the constraint algebra Eq.~(\ref{Jalg}),
are needed to evaluate an N-point amplitude.  Consider, for example,
the coupling of three massless vector particles, given by ${\cal A}_3 =
\langle W_D(k_3,\xi_3)|V_D(k_2,\xi_2; 1)|W_D(k_1,\xi_1)\rangle$.
One finds
\be\label{A3}
{\cal A}_3=i2\sqrt{2}\Bigl[(k_1\cdot\xi_3)(\xi_2\cdot\xi_1)+
(k_2\cdot\xi_1)(\xi_3\cdot\xi_2)+(k_3\cdot\xi_2)(\xi_1\cdot\xi_3)
-3(\xi_1\cdot k_2) (\xi_2\cdot k_3) (\xi_3\cdot k_1)\Bigr]\ .
\ee
The first three terms are precisely the expected Yang-Mills coupling;
the last term represents a non-linear correction to the Yang-Mills
action which is higher-order in the string tension, and therefore
is suppressed at energies far below the Planck scale. The non-linear
term also appears in the three-vector coupling in the bosonic
string, where it has coefficient $+1$ instead of $-3$; in the
superstring no such term appears in the three-point coupling
(though string correction terms do appear in higher-point functions).

One can calculate higher-point amplitudes in a similar way.  While
the details are not illuminating, the main features of these
amplitudes are easily understood.  Because the contribution from
the ${\rm e}^{ik\cdot X}$ pieces essentially factorizes, it is
clear that one will obtain gamma-function factors similar to
those that appear in the Veneziano amplitude for the bosonic
string.  The remaining part of the vertices can only contribute
factors which are polynomial in the momenta.  Thus, the
fractional string amplitudes
have the soft high-energy Regge behavior characteristic
of bosonic and superstring amplitudes.

\vglue 0.6cm
%===============================================================
{\elevenbf\noindent 5. Determination of the critical central charge}
\vglue 0.4cm
Spurious state decoupling by itself does not imply that tree
amplitudes are unitary.  One must also prove that there are no
negative-norm physical states for the specific representation of
the constraint algebra under consideration.  In the bosonic and
superstrings, for representations with one time-like (space-time)
dimension, a non-negative physical state space occurs up to a
maximum value of the central charge.  As one passes through
this critical value of the central charge the norm of
some physical states change
sign, implying that at the critical central charge
there are extra null states.  We can
check for the existence of a critical central charge in a
representation-independent way by searching for the occurence of
extra sets of zero-norm physical states.

Since a null physical state is spurious, consider the simplest (D-sector)
spurious state $\ket{\psi_1}=G_{-1/3}\ket{\widetilde\psi}$, where
$\widetilde\psi$ is an arbitrary state.  The physical state
conditions Eq.~(\ref{psc}) are satisfied by $\psi_1$ only when
the intercept $v=1/3$, irrespective of the value of the central charge.
This is the value of the D-sector intercept determined above by the
requirement of spurious state decoupling;  when $\widetilde\psi
= {\rm e}^{ik\cdot X}$, $\psi_1$ is just the longitudinal
component of the massless vector state in Eq.~(\ref{vectvert}).

To determine the critical central charge we must consider the
more complicated general D-sector spurious state
\be\label{psi2}
\ket{\psi_2}=\left(G_{-4/3}+\alpha G_{-1/3}L_{-1}
+\beta G_{-1}G_{-1/3}\right)\ket{\widetilde\psi}\ .
\ee
Applying the physical state conditions, one finds a null state for
specific values of $\alpha$ and $\beta$ only when $c=10$.$^{1,10}$
One can check in the $c=5$ sub-critical representation of the
last section that the lowest-mass state of the form in Eq.~(\ref{psi2})
(taking $\widetilde\psi={\rm e}^{ik\cdot X}$) is in fact a
positive-norm state, as required for tree-level unitarity.

One can also search for sets of S-module null physical states.
In particular, for the general spurious states of the form
$\ket{\psi_S}=G_{-2/3}\ket{\widetilde\psi}$, the physical state
conditions imply a quadratic relation between the central
charge and the S-sector intercept $v_S$.  For $c=10$, the
solutions are $v_S=1/3$ or $-1$.  The first possibility gives rise
to tachyons, whereas the second describes only massive states.
\vglue 0.6cm
%===============================================================
{\elevenbf\noindent 6. Discussion and outlook}
\vglue 0.4cm
The tree-level considerations discussed so far leave us with
a certain amount of arbitrariness in constructing spin-4/3
fractional superstrings.  In particular, we are free to include or
not S-sector states;
we can couple left- and right-moving theories at will on the
world-sheet in type II and heterotic constructions; and the
choice of CFT representation of the spin-4/3 FSC algebra is
constrained by tree unitarity only to have central charge less
than or equal to its critical value $c=10$.  The inclusion of
string loop amplitudes should remove much of this arbitrariness.
As is the case with the bosonic and superstrings, one expects that loop
amplitudes will only be consistent at the critical central charge,
and modular invariance will determine which left- and right-moving
sectors, at which values of their intercepts, can be consistently
coupled together.

One difficulty in constructing a critical ($c=10$) representation
of the FSC algebra is its non-linearity. The fact that the FSC
structure constant $\lambda$ is a non-linear function of the
central charge $c$, Eq.~(\ref{lamb}), means that the FSC algebra as a whole
is non-linear: the tensor product of two representations
of this algebra is not itself a representation.
In particular, the tensor product of two copies of the $c=5$
representation described above will not make a $c=10$ representation
of the FSC algebra.  One can, however, construct higher-$c$
representations from a given representation by turning on a
background charge for one of the $X^\mu(z)$ coordinate boson
fields, corresponding to a linear dilaton background
field in space-time.$^{5}$

Once given a $c=10$ representation of the FSC algebra, the
construction of loop amplitudes may still not be an easy task.
One can imagine ``sewing'' tree amplitudes in the ``old covariant''
formalism described above to form one-loop amplitudes by a suitable
generalization of the sewing procedure for the bosonic
string.$^{11}$  Such an amplitude would not only have to
be unitary, but also modular invariant.  An indication that this
may not be too much to expect may be the existence of the
modular-invariant fractional string partition functions proposed
in Ref.~2.
At higher loops it seems likely that a clearer understanding
of the ``fractional moduli'' describing the sewing of tree
amplitudes will be necessary.  This is essentially the question
of what is the local world-sheet symmetry underlying the FSC
constraint algebra.  Though the form of the FSC algebra provides a
rigid guide to such a symmetry, its identification remains an
open question.

Finally, what are the space-time features of critical fractional
superstrings?  This, also, is an open question, since its answer
depends largely on the resolution of the world-sheet issues
outlined above.  So far, only hints of possibly new space-time
structures have been gleaned from the fractional superstring
partition function.$^{3,4,12,13,14}$
\vglue 0.6cm
%===============================================================
{\elevenbf\noindent Acknowledgements}
\vglue 0.4cm
This talk reports on joint work with S.-H. Henry Tye.$^{5}$
It is a pleasure to thank K.~Dienes and J.~Grochocinski
for useful discussions and comments.  I am grateful
to the Center for Theoretical Physics at M.I.T., and to
B.~Zwiebach in particular, for their hospitality.  This work
was supported in part by the National Science Foundation.
\vglue 0.5cm
%===============================================================
{\elevenbf\noindent References}
\vglue 0.4cm


\begin{thebibliography}{99}
\bibitem{ALT}P.C.~Argyres, A.~LeClair and S.-H.H.~Tye,
{\elevenit Phys. Lett.} {\elevenbf 253B} (1991) 306.
\bibitem{AT}P.C.~Argyres and S.-H.H.~Tye, {\elevenit Phys. Rev. Lett.}
{\elevenbf 67} (1991) 3339.
\bibitem{DT}K.R.~Dienes and S.-H.H.~Tye, {\elevenit Nucl. Phys.}
{\elevenbf B376} (1992) 297.
\bibitem{ADT}P.C.~Argyres, K.R.~Dienes and S.-H.H.~Tye,
{\elevenit New Jacobi-like Identities for Parafermion Characters},
Cornell/McGill preprint CLNS 91/1113, McGill/91-37, to appear in
{\elevenit Comm. Math. Phys.}
\bibitem{AT2}P.C.~Argyres and S.-H.H.~Tye, {\elevenit Tree Scattering
Amplitudes for the K=4 Fractional Superstring}, Cornell
preprint CLNS 92/1176, to appear.
\bibitem{ZFpara}A.B.~Zamolodchikov and V.A.~Fateev,
{\elevenit Sov. Phys. J.E.T.P.} {\elevenbf 62} (1985) 215;
{\elevenit Sov. Phys. J.E.T.P.} {\elevenbf 63} (1986) 913.
\bibitem{FZft}A.B.~Zamolodchikov and V.A.~Fateev,
{\elevenit Theor. Math.  Phys.} {\elevenbf 71} (1987) 451.
\bibitem{GoS}P.~Goddard and A.~Schwimmer, {\elevenit Phys. Lett.}
{\elevenbf 206B} (1988) 62.
\bibitem{GSW}M.B.~Green, J.H.~Schwarz and E.~Witten,
{\elevenit Superstring Theory,} Cambridge University Press (1987).
\bibitem{ALyT}P.C.~Argyres, E.~Lyman and S.-H.H.~Tye, {\elevenit
Phys. Rev.} {\elevenbf D46} (1992) 4533.
\bibitem{BrO}L.~Brink and D.I.~Olive, {\elevenit Nucl. Phys.}
{\elevenbf B56} (1973) 253; {\elevenit Nucl. Phys.} {\elevenbf B58}
(1973) 237.
\bibitem{FL}P.~Frampton and J.~Liu, {\elevenit Phys. Rev. Lett.}
{\elevenbf 70} (1993) 130.
\bibitem{CR}G.B.~Cleaver and P.J.~Rosenthal, {\elevenit Aspects of
Fractional Superstrings,} Caltech preprint CALT-68-1756.
\bibitem{AD}P.C.~Argyres and K.R.~Dienes, {\elevenit On the Massive
Sectors and Internal Projections of the Fractional Superstring,}
Cornell/McGill preprint CLNS 92/1168, McGill/92-41, to appear.
\end{thebibliography}
\end{document}